\documentclass{article}
\usepackage{orcidlink}
\usepackage{amsmath,amssymb,amsfonts}
\usepackage[margin=4cm]{geometry}
\usepackage{algpseudocode}
\usepackage{algorithm}
\usepackage{algorithmicx}
\usepackage{arxiv}
\usepackage[utf8]{inputenc} % allow utf-8 input
\usepackage[T1]{fontenc}    % use 8-bit T1 fonts
\usepackage{hyperref}       % hyperlinks
\usepackage{url}            % simple URL typesetting
\usepackage{graphicx}
\usepackage{plantuml}
\usepackage{verbatim}
\usepackage{multirow}
\usepackage{booktabs}       % professional-quality tables
\usepackage{flafter}		% for formatting tables correctly (tables stuck at top / not under section  issue)
\usepackage{longtable}
\usepackage{acronym}
\usepackage[acronym]{glossaries}
\usepackage{diagbox}

\newacronym{GAN}{GAN}{Generative Adversarial Networks}
\newacronym{VAE}{VAE}{Variational Autoencoder}
\newacronym{ML}{ML}{Machine Learning}
\newacronym{AI}{AI}{Artificial Intelligence}
\newacronym{NLP}{NLP}{Natural Language Processing}
\newacronym{PII}{PII}{Personally Identifiable Information}
\newacronym{LDP}{LDP}{Local Differential Privacy}
\newacronym{CDP}{CDP}{Central Differential Privacy}
\newacronym{DP}{DP}{Differential Privacy}
\newacronym{DP-SGD}{DP-SGD}{Differentially Private Stochastic Gradient Descent}
\newacronym{BFSI}{BFSI}{Banking, Financial Services, and Insurance}
\newacronym{GDPR}{GDPR}{General Data Protection Regulation}
\newacronym{CCPA}{CCPA}{California Consumer Privacy Act}
\newacronym{CFE}{CFE}{Counterfactual Fairness Evaluation}

\newacronym{EHR}{EHR}{Electronic Health Records}
\graphicspath{ {./images/} }
\title{ Optimizing the Privacy-Utility Balance using Synthetic Data and Configurable Perturbation Pipelines }

\date{Apr 10, 2025}	% Here you can change the date presented in the paper title
\author{ 
    \href{https://orcid.org/0000-0002-9064-3362}
    {\includegraphics[scale=0.06]{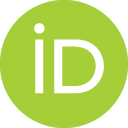}\hspace{1mm}Anantha Sharma}\\
    Head of AI - Architecture \& Strategy \\
    \And
    \href{https://orcid.org/0009-0005-6532-9389}
    {\includegraphics[scale=0.06]{orcid}\hspace{1mm}Swetha Devabhaktuni}\\
    Head of Data \& Analytics - North America\\
    \And
    \href{https://orcid.org/0009-0009-0380-9343}
    {\includegraphics[scale=0.06]{orcid}\hspace{1mm}Eklove Mohan}\\
    CTO Office - North America\\
}
\renewcommand{\headeright}{}
\renewcommand{\undertitle}{ }

\begin{document}

\maketitle

\begin{abstract}
    This paper explores the strategic use of modern synthetic data generation and advanced data perturbation techniques to enhance security, 
    maintain analytical utility, and improve operational efficiency when managing large datasets, with a particular focus on the Banking, 
    Financial Services, and Insurance (BFSI) sector. We contrast these advanced methods encompassing generative models like GANs, 
    sophisticated context-aware PII transformation, configurable statistical perturbation, and differential privacy with traditional anonymization approaches. 
    The goal is to create realistic, privacy-preserving datasets that retain high utility for complex machine learning tasks and analytics, a critical need in the data-sensitive industries like BFSI, Healthcare, Retail, and Telecommunications. We discuss how these modern techniques potentially offer significant improvements in balancing privacy preservation while maintaining data utility compared to older methods. Furthermore, we examine the potential for operational gains, such as reduced overhead and accelerated analytics, by using these privacy-enhanced datasets. We also explore key use cases where these methods can mitigate regulatory risks (e.g., GDPR \cite{ep2016a} ) and enable scalable, data-driven innovation without compromising sensitive customer information
\end{abstract}

\section{Introduction}

The Banking, Financial Services, and Insurance (BFSI) sector operates on vast volumes of highly sensitive customer data, creating an enduring tension between the drive for data‑driven insights and the imperative to comply with strict privacy and security regulations such as GDPR \cite{ep2016a} and CCPA \cite{ccpa}. Traditional anonymization methods like masking, aggregation, k‑anonymity, L‑diversity, and T‑closeness often degrade data quality to the point where sophisticated analytics, fraud detection, risk modeling, and machine learning applications suffer significant performance drops. Moreover, these legacy approaches can remain vulnerable to linkage and inference attacks, undermining both privacy guarantees and competitive innovation in financial institutions. The need for advanced techniques that can create privacy-preserving datasets without sacrificing analytical utility is paramount.

In response, advanced techniques for creating privacy-preserving datasets have emerged, broadly categorized as purely synthetic data generation and advanced data perturbation. \textbf{Purely synthetic data}, often created using deep generative models (like GANs ), aims to capture the statistical patterns of real data without any one-to-one mapping to real individuals. \textbf{Advanced data perturbation} applies carefully calibrated noise, transformations, and privacy-enhancing techniques like differential privacy  to original datasets, seeking to obscure sensitive information while retaining analytical value. These methods can include context-aware transformations, where the nature of the data and its intended use inform the perturbation strategy, ensuring that the resulting dataset remains useful for specific tasks. However, the challenge remains to balance privacy and utility effectively. Traditional methods often fail to provide sufficient privacy guarantees or result in datasets that are too noisy for practical use. In contrast, modern techniques like generative models and advanced perturbation strategies can offer a more nuanced approach, allowing for the generation of high-fidelity synthetic datasets that maintain the statistical properties of the original data while ensuring compliance with privacy regulations.

The BFSI sector is particularly well-suited for these advanced techniques due to its data-rich environment and the critical need for robust analytics. Financial institutions are increasingly turning to machine learning and AI to drive insights, improve customer experiences, and enhance operational efficiency. However, the sensitive nature of the data involved necessitates a careful approach to privacy and security. This paper provides a comparative perspective on these modern approaches versus traditional methods. We explore how configurable pipelines combining statistical perturbation, differential privacy, context-aware PII transformation, and generative models can potentially overcome the limitations of older techniques. We specifically focus on the utility and applicability of these methods within the BFSI domain, examining use cases where maintaining data fidelity for complex tasks is paramount while adhering to strict compliance requirements. The objective is to evaluate how these advanced strategies can better balance the critical needs for data privacy, analytical utility, and operational efficiency in financial institutions.

\section{ Background}

The increasing reliance on large-scale datasets has heightened concerns regarding privacy and operational efficiency. To address these challenges, various anonymization techniques have been developed, including differential privacy, masking, perturbation, and others.

Modern privacy-enhancing solutions often combine these techniques within configurable frameworks, allowing practitioners to tailor the approach to specific data types and use case requirements.

\section{Traditional Anonymization}
Techniques like k-anonymity \cite{k_anonimity}, L-diversity \cite{l_diversity}, T-closeness \cite{t_closeness}, aggregation, and basic masking/tokenization aim to prevent re-identification by generalizing or suppressing data. 

However, they often struggle with high-dimensional data which can significantly reduce data utility (especially for ML), and may still be vulnerable to linkage or inference attacks \cite{trouble_with_kanon_and_tcloseness}.

\subsection{Masking} Masking is a technique that replaces sensitive data with non-sensitive equivalents, such as replacing names with asterisks or pseudonyms. While this can effectively obscure direct identifiers, it may not sufficiently protect against linkage attacks if the masked data retains enough structure or context. For example, if a dataset contains masked names but retains other identifying attributes (e.g., birth dates, addresses), an adversary may still be able to re-identify individuals by correlating the masked data with external datasets. 

Simple masking, while easy, destroys analytical value in the masked fields.

\textbf{Example}:

    \begin{table}[ht]
        \centering
        \caption{Masking Example}
        \begin{tabular}{|c|c|c|}
            \hline
            \textbf{Original Text} & \textbf{Masked Text} & \textbf{Notes} \\
            \hline
            555.192.9277 & 555.XXX.XXXX & Phone number masking (using regex) \\
            \hline
            5423 3428 2372 9072 & 5XX3 XXXX XXXX 9072 & Credit card masking (using regex) \\
            \hline
            123 Any Street, Canada City, Canada & XXX Any Street, Canada City, Canada & Address masking (using NER \& regex) \\
            \hline
        \end{tabular}
        \label{tab:example_masking}
    \end{table}

    Our work \cite{deshmukh2023lifepiipii} and \cite{tableguard} proposes a more advanced approach to masking, which involves context-aware transformations that consider the nature of the data and its intended use. This method aims to obscure sensitive information while preserving the statistical properties of the dataset, making it more suitable for complex analytics and machine learning tasks.

\subsection{Statistical Perturbation}

Statistical perturbation is a cornerstone technique in generating synthetic data for enhancing privacy, particularly within frameworks like differential privacy. This approach modifies original data values through controlled distortions to prevent identification of individual records. 

\textbf{Noise Addition} is a fundamental method which involves adding random noise to numerical attributes. The choice of noise distribution hinges on the sensitivity  $ \delta $ of the data and the privacy budget $\epsilon$. For instance, Laplace noise is often preferred for its simplicity in guaranteeing differential privacy when applied to sums or means. However, excessive noise can degrade signal fidelity, necessitating a careful balance between privacy strength and utility.

\subsection{Multiplicative Perturbation} 

This method is particularly useful for preserving the relative relationships between data points while introducing uncertainty, such as in financial datasets where ratios or proportions matter. 
It requires careful calibration to avoid distorting the underlying data distribution. 

$ y_i = x_i \times k_i $
where $ y_i $ is the perturbed value, $ x_i $ is the original value, and $ k_i $ is a random factor drawn from a specified distribution, effectively scaling the data.

For example, if income values are perturbed using a multiplicative factor drawn from a uniform distribution between $0.8$ and $1.2$, 

The resulting dataset will still reflect the original income distribution while adding a layer of privacy.

Here, $ k_i $ is typically drawn from a carefully calibrated distribution to balance privacy and utility. 
Common choices for $ k_i $ include uniform or gamma distributions. 
For example, if $ k_i \sim Uniform(a, b) $, the perturbation factor is selected uniformly between $ a $ and $ b $. 
To ensure differential privacy guarantees, the choice of $ k_i $ must be such that the sensitivity of the data is appropriately bounded. 
This method is particularly effective for \textbf{positive-valued data} where scaling maintains proportional relationships.

\textbf{Example}:

    \begin{table}[ht]
        \centering
        \caption{Multiplicative Perturbation Example}
        \begin{tabular}{|c|c|c|}
            \hline
            \textbf{Original Value} & \textbf{Perturbed Value} & \textbf{Notes} \\
            \hline
            1000 & 1200 & Income value scaled by 1.2 \\
            \hline
            2000 & 1600 & Income value scaled by 0.8 \\
            \hline
        \end{tabular}
        \label{tab:example_multiplicative_perturbation}
    \end{table}

\textbf{Extending with additive noise}
Multiplicative perturbation can be combined with additive noise. For instance:

$ y_i = x_i \times k_i + \text{Lap}\left(\frac{\Delta f}{\epsilon}\right) $

Here, $ \text{Lap}(\cdot) $ denotes Laplace noise scaled by the sensitivity $ \Delta f $ and privacy parameter $ \epsilon $. 
This hybrid approach combines scaling with noise addition to enhance both privacy and utility.

\textbf{Example}:

    \begin{table}[ht]
        \centering
        \caption{Multiplicative Perturbation with Additive Noise Example}
        \begin{tabular}{|c|c|c|}
            \hline
            \textbf{Original Value} & \textbf{Perturbed Value} & \textbf{Notes} \\
            \hline
            1000 & 1200 + Lap(0, 0.1) $ \approxeq $ ( 1200.07, 1199.98 ) & Income value scaled by 1.2 before adding Laplace noise \\
            \hline
            2000 & 1600 + Lap(0, 0.1) $ \approxeq $ ( 1600.12, 1599.85 ) & Income value scaled by 0.8 before adding Laplace noise \\
            \hline
        \end{tabular}
        \label{tab:example_multiplicative_perturbation_with_noise}
    \end{table}

\subsection{Binning} Binning or discretization is a process in which continuous variables are transformed into discrete intervals (e.g., age groups, income brackets, or credit score ranges). This process reduces data sensitivity by limiting the granularity of values, thus mitigating the risk of disclosure. However, this reduction in granularity may lead to a loss of resolution and can negatively impact downstream analyses if the bin edges are not chosen carefully.

For example, consider a dataset of credit scores that is segmented into the following ranges:
\begin{itemize}
    \item \textbf{Poor:} 300--579
    \item \textbf{Fair:} 580--669
    \item \textbf{Good:} 670--739
    \item \textbf{Very Good:} 740--799
    \item \textbf{Excellent:} 800--850
\end{itemize}

If a credit score of 669 (which is at the upper edge of the ``Fair'' category) is perturbed due to noise or misclassification during the binning process, it could be erroneously placed in the ``Good'' category. Such misclassification may result in improper risk assessment or erroneous decisions regarding interest rates or loan eligibility.

Binning is often combined with noise addition, for instance, adding Laplace noise to further enhance privacy without completely sacrificing data resolution. 
However, if the bin boundaries do not align with natural clusters in the data (as in age groups or income distributions), this may also lead to sparse bins, unstable statistical estimates, and even re-identification risks \cite{reidentificationofanonymizeddata}.

\begin{table}[htbp]
    \centering
    \caption{Examples of Binning and Potential Misclassification}
    \label{tab:binning-examples}
    \begin{tabular}{@{}lcccc@{}}
        \toprule
        \textbf{Variable} & \textbf{Continuous Value} & \textbf{Noise/Perturbation} & \textbf{Binned Interval} & \textbf{Remarks} \\ \midrule
        Age & 29 & None & 20--29 & Proper classification \\
        Age & 29 & +2 (perturbed) & 30--39 & Within acceptable range \\ \midrule
        Credit Score & 669 & None & 580--669 (\textit{Fair}) & Proper classification \\
        Credit Score & 669 & +2 (perturbed) & 670--739 (\textit{Good}) &  \textcolor{red}{Outside acceptable range} \\ \midrule
        Income & \$45,000 & None & \$40K--\$50K & Appropriate binning \\
        Income & \$45,000 & -\$3,000 (perturbed) & \$40K--\$50K & Within acceptable range \\ \bottomrule
    \end{tabular}
\end{table}

In summary, while binning can significantly reduce the risk of privacy breaches by limiting the resolution of data, care must be taken in selecting bin edges. Inappropriate binning, especially when combined with noise addition, can lead to unintended consequences such as misclassification of critical values (e.g., a credit score moving from ``Fair'' to ``Good''). This trade-off must be carefully balanced in any data anonymization strategy.

\subsection{Clipping} is a preprocessing technique that constrains continuous numerical data within a predefined range (e.g., capping values at certain quantiles). The sensitivity of each data point is bounded by “clipping” extreme values, which can make subsequent noise addition (as in differential privacy) more effective. For instance, clipping income values to 
$
[\mu - 3\sigma,\, \mu + 3\sigma]
$
ensures that extreme outliers do not disproportionately influence aggregate statistics.

\begin{figure}
    \centering
    \includegraphics[width=0.8\textwidth]{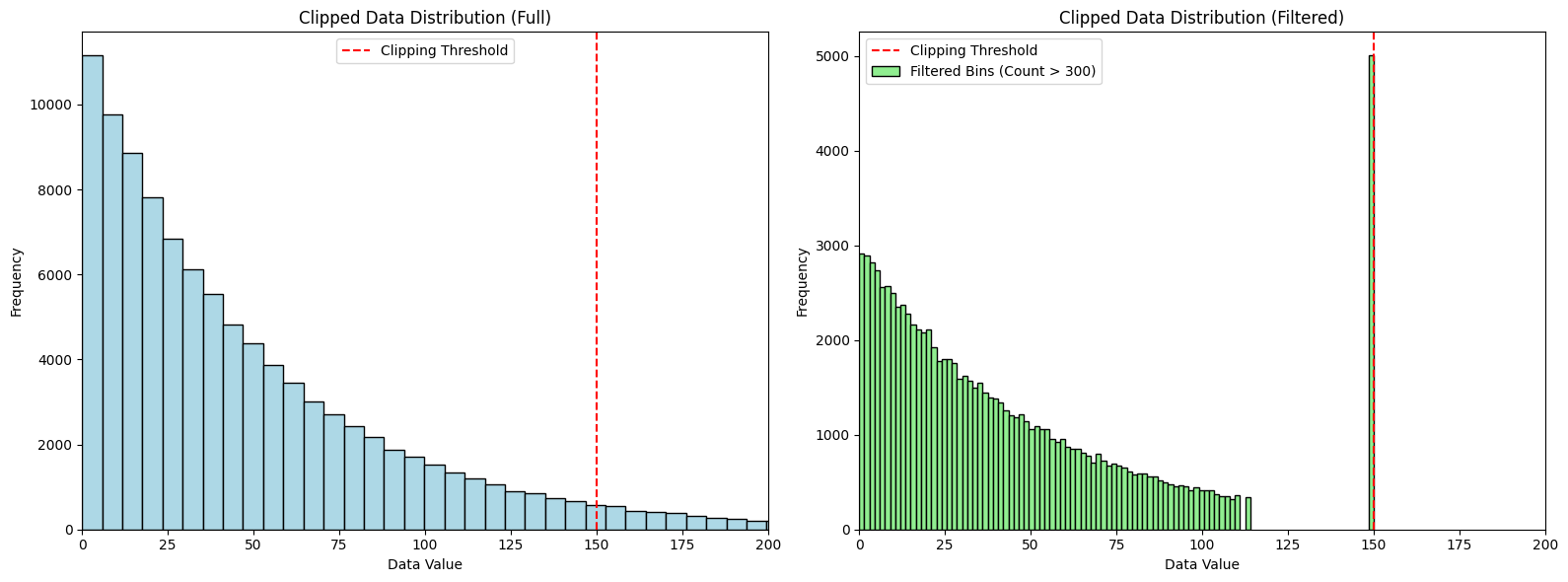}
    \caption{Clipping: Original Data vs. Clipped Data}
    \label{fig:clipping_example}
\end{figure}

\begin{itemize}

    \item \textbf{Residual Risk from Dense Concentrations:} Even after clipping, if many records lie near the clipping threshold, an adversary may focus on these “bunched” values to perform inference attacks. For example, if most incomes are clipped at the upper bound, an attacker might correlate auxiliary data to deduce which individuals originally were high earners.
    \item \textbf{Increased Risk of Inference Attacks:} Clipping can create artificial clusters of similar values, making it easier for adversaries to infer sensitive information. For example, if a dataset contains ages that are clipped at 30 and 60, an attacker may deduce that individuals in the 30-60 age range are more likely to be in a specific demographic group.
    \item \textbf{Loss of Information:} Clipping can lead to a loss of information, especially in datasets with heavy-tailed distributions. For instance, if a dataset contains income values that are heavily skewed, clipping may remove valuable information about the tail behavior, which could be critical for certain analyses (e.g., risk assessment).
    \item \textbf{Bias in Statistical Analysis:} Clipping can introduce bias in statistical analyses, especially if the clipped values are not representative of the underlying distribution. For example, if a dataset contains income values that are clipped at the upper bound, the resulting dataset may not accurately reflect the true distribution of incomes, leading to biased estimates of mean or median income.
    \item \textbf{Reduced Variability:} Clipping can reduce the variability of the data, which may lead to biased estimates in statistical analyses. For example, if a dataset contains income values that are clipped at the upper bound, the resulting dataset may not accurately reflect the true distribution of incomes, leading to biased estimates of mean or median income.
    \item \textbf{Increased Risk of Re-identification:} Clipping can inadvertently create unique or near-unique records, especially in high-dimensional datasets. For example, if a dataset contains multiple attributes (e.g., age, income, and education level) that are all clipped, the combination of these clipped values may still yield unique records that can be re-identified using external auxiliary data.
    \item \textbf{Increased Complexity in Data Analysis:} Clipping can complicate data analysis, especially when dealing with complex datasets. For example, if a dataset contains multiple attributes that are all clipped, the resulting dataset may be more difficult to analyze and interpret, leading to potential errors in data analysis.
    \item \textbf{Vulnerability in High-Dimensional Data:} In datasets with many continuous attributes, clipping is typically applied independently to each field. Consequently, the combination of clipped values may still yield unique or near-unique records. Such multi-dimensional uniqueness, when combined with external auxiliary data, can facilitate re-identification.
\end{itemize}

 For example, in studies on the de-identification of Electronic Health Records (EHRs), researchers have applied clipping to laboratory test results (e.g., blood sugar levels) to limit the influence of measurement errors or outliers. However, record linkage attacks have shown that even with clipped values, the combination of multiple clipped test results can render individual patients re-identifiable especially if the underlying distribution is heavy-tailed \cite{riskofreidentification} (see \ref{fig:clipping_example} ).

Clipping is not a foolproof method for ensuring anonymity; however, it is useful for reducing the impact of outliers and bounding sensitivity. Adversaries can often leverage the residual structure and combine clipped data with external information to re-identify or infer sensitive details.

These methods collectively address the privacy-utility trade-off by introducing controlled uncertainty while preserving essential statistical properties. However, their effectiveness depends on precise parameter tuning and domain-specific considerations. For example, multiplicative perturbation might be preferred in healthcare data to maintain proportional relationships between features, whereas binning could be more suitable for categorical transformations in financial datasets.

\begin{table}[htbp]
    \centering
    \caption{Noise Comparison Table (Base Value = 1200, Noise Level = 0.1)}
    \begin{tabular}{|l|l|l|l|l|}
        \hline
        \textbf{Noise Type} & \textbf{Distribution} & \textbf{Sampled Noise (Example)} & \textbf{Final Value (1200 + noise)}  \\ \hline
        Laplace    & $ \text{Lap}(0, 0.1) $            & $+0.07$, $-0.12$             & $1200.07$, $1199.88$   \\ \hline
        Gaussian   & $ \mathcal{N}(0, 0.01) $         & $+0.05$, $-0.15$             & $1200.05$, $1199.85$  \\ \hline
        Uniform    & $ \mathcal{U}(-0.1, 0.1) $       & $+0.08$, $-0.03$             & $1200.08$, $1199.97$  \\ \hline
        Cauchy     & $ \text{Cauchy}(0, 0.1) $         & $+0.2$, $-1.5$ (extreme)       & $1200.2$, $1198.5$   \\ \hline
        Cholesky & $ L \cdot z $ (scalar: $ \sigma = 0.1 $) & $+0.06$                        & $1200.06$  \\ \hline
    \end{tabular}
\end{table}

\subsection{Differential Privacy (DP)}
Differential Privacy (DP) is a formal framework that provides provable privacy guarantees. It ensures that the outcome of any analysis changes only minimally when any single individual's data is added or removed. This is typically achieved by adding carefully calibrated noise based on the function's sensitivity and a privacy budget parameter, usually denoted by $\epsilon$ (and sometimes $\delta$ for approximate DP).

\paragraph{Usage:}
\begin{itemize}
    \item \textbf{Count Queries:} When counting occurrences of specific values or patterns in a dataset, noise (e.g., Laplace or Gaussian noise) is added to hide the true count.
    \item \textbf{Sum Queries:} The Laplace mechanism can be applied to the sum of values, ensuring that the released total is differentially private.
    \item \textbf{Mean Queries:} Add noise to the mean of values and make it differentially private.
    \item \textbf{Histogram Queries:} Differential privacy is achieved by adding noise to the counts in each bin, so that the overall histogram does not reveal individual contributions.
\end{itemize}

\paragraph{Implementations of Differential Privacy:}
\begin{itemize}
    \item \textbf{Local Differential Privacy (LDP):} In LDP, each individual perturbs their own data before sending it to a central server. This ensures that the server never sees the raw data, providing strong privacy guarantees at the cost of higher noise levels. \cite{apple2023learningwithprivacyatscale}
    \item \textbf{Central Differential Privacy (CDP):} Here, raw data is collected by a trusted curator who then adds noise to the aggregated query results. This model typically allows for more accurate analyses since noise can be optimized globally. 
    \item \textbf{Approximate Differential Privacy:} This is a relaxation of strict DP that permits a small probability, $\delta$, that the privacy guarantee may be violated. This trade-off can improve utility in cases where strict DP is too restrictive.
    \item \textbf{Differentially Private Stochastic Gradient Descent (DP-SGD):} DP-SGD is an algorithmic modification of standard stochastic gradient descent where noise is added to the gradient updates during training. This enables machine learning models to be trained on sensitive data while providing formal privacy guarantees.
\end{itemize}

Differential privacy has seen practical applications across various domains. 

For instance, the U.S. Census Bureau adopted differential privacy for the 2020 Census to protect respondent confidentiality \cite{Dwork2014, uscensus2018}. In healthcare, DP mechanisms enable the release of aggregate statistics while mitigating re-identification risks. Moreover, DP-SGD has been widely used in training deep learning models to ensure that the contribution of any single individual's data does not significantly affect the outcome \cite{mcmahan2018learningdifferentiallyprivaterecurrent}.

\section{Advanced Data Generation Strategies}

Modern approaches offer more sophisticated ways to create privacy-preserving datasets compared to traditional anonymization.
These methods can be broadly categorized into two main strategies: \textbf{purely synthetic data generation} and \textbf{advanced data perturbation}. Each approach has its own strengths and weaknesses, and the choice between them often depends on the specific use case, regulatory requirements, and desired balance between privacy and utility.

\begin{figure}[htbp]
    \centering
    \includegraphics[width=0.95\textwidth]{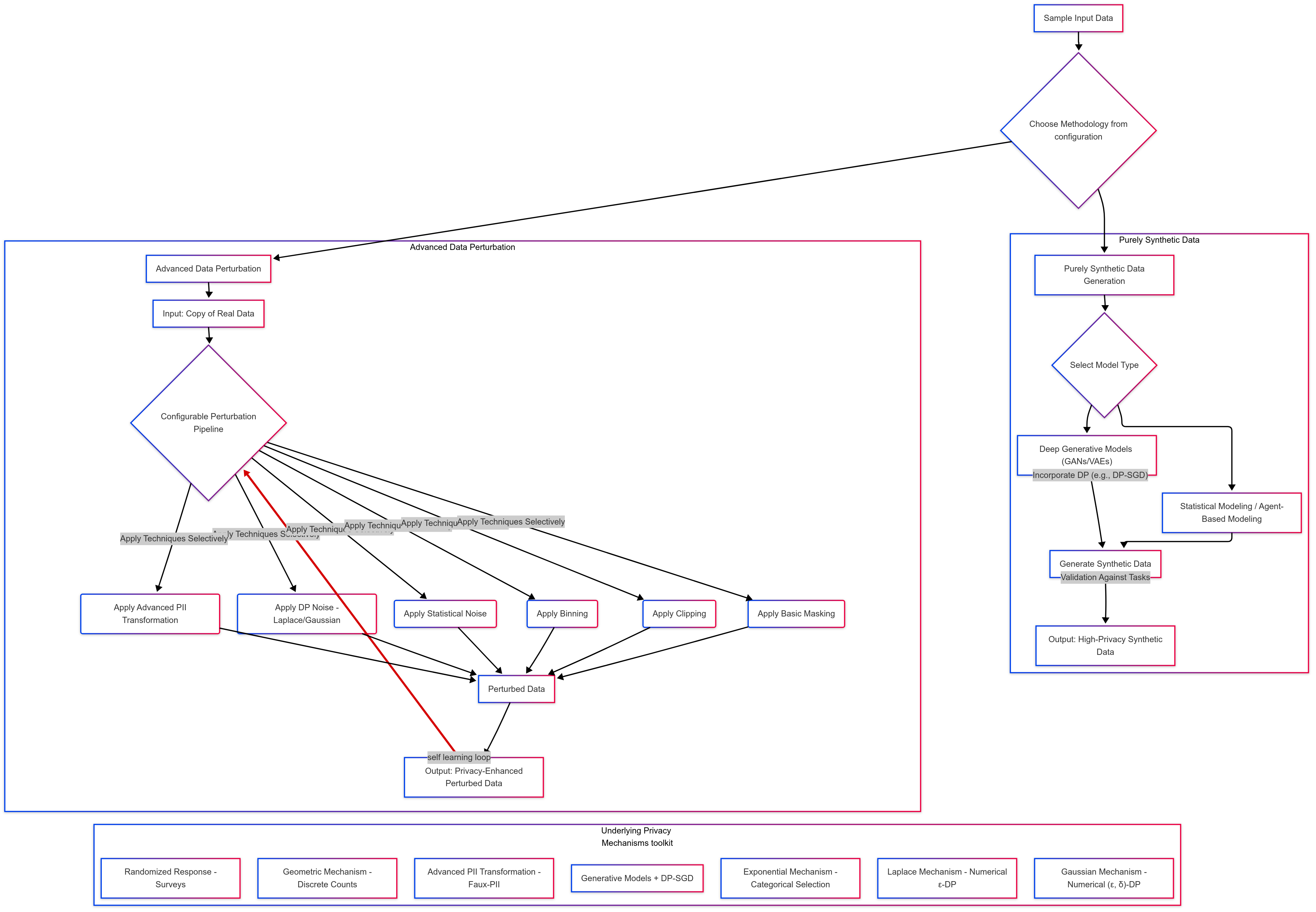}
    \caption{Highlevel view of advanced Differential Privacy}
    \label{fig:advanced_differential_privacy}
\end{figure}

\subsection{Purely Synthetic Data}

Ensuring the synthetic data accurately reflects all relevant statistical properties and complex correlations of the real data (high fidelity) can be difficult. Generative models can be computationally expensive to train and may suffer from issues like mode collapse (failing to capture all variations). Incorporating DP during training adds complexity but provides formal privacy guarantees. Validation against real-world tasks is crucial and provide the highest level of privacy as there's no direct link to real individuals.

Methods include Statistical modeling (e.g., fitting distributions and sampling), Agent-Based Modeling, and Deep Generative Models (GANs, VAEs) \cite{sdinhealthcare} are common. Generative models can capture complex, non-linear relationships and dependencies within the data.

\subsection{Advanced Data Perturbation}

This applies transformations directly to copies of real datasets by utilizing configurable sequences of techniques described earlier (DP noise, statistical noise, binning, clipping, masking, advanced PII transformation). Modern platforms allow applying different techniques with different parameters to specific columns based on their type and sensitivity. For instance, when perturbing credit scores, rules can prevent noise addition from illogically crossing critical thresholds (e.g., $720$). Similarly, PII transformation aims to replace names with realistic fake names, not just asterisks. This context can be manually defined or potentially informed by AI/LLM analysis of column data.

These combined techniques can often achieve a good balance between privacy and utility by directly modifying real data patterns rather than recreating them from scratch. This approach can be computationally less intensive than training complex generative models and allow fine-grained control over the perturbation process per attribute.

However, this still requires careful calibration to avoid excessive utility loss. Privacy guarantees are typically linked to specific techniques used (e.g., the $\epsilon $ or $ \delta $ values if DP is applied) rather than the entire dataset being fundamentally disconnected from real individuals. Vulnerability to reconstruction attacks needs consideration depending on the methods used.

\begin{itemize}
    \item Pure Differential privacy is where Laplace, Exponential, etc. techniques use a single parameter $ \epsilon $.
    \item Approximate Differential privacy (Gaussian mechanism) uses two parameters ($\epsilon$, $\delta$).
\end{itemize}

\subsection{Balancing Privacy and Utility}
Both approaches face the fundamental trade-off between privacy protection and data utility.

\begin{itemize}
    \item \textbf{Configurability:} Advanced platforms allow users to tune parameters (e.g., noise levels, DP epsilon, bin sizes, clipping bounds) to navigate this trade-off
    \item \textbf{Statistical Fidelity:} Comparing distributions (e.g., using KS/Chi2 tests), correlations, and basic statistics between the original and processed data
    \item \textbf{Machine Learning Utility:} Evaluating the performance difference of ML models (e.g., classification accuracy, AUC) trained on the original versus the privacy-enhanced data
    \item \textbf{Domain Knowledge:} Selecting appropriate techniques and parameters requires understanding the data and the downstream task. AI-assisted analysis  might help by identifying sensitive fields or suggesting relevant constraints for perturbation
\end{itemize}

\section{Privacy-Preserving Mechanisms}

Differential privacy and synthetic data generation rely on several key mechanisms to balance privacy and utility. This section outlines the primary mechanisms and their applications in privacy-preserving data publishing.
\subsection{Geometric Mechanism}
The geometric mechanism is designed to preserve privacy in count queries by adding noise sampled from a geometric distribution. This mechanism is particularly effective when dealing with discrete data and queries that have a well-defined, bounded global sensitivity. 

For a query function $ f: \mathcal{D} \rightarrow \mathbb{Z} $, the noisy output is computed as:
$
\tilde{f}(D) = f(D) + \eta,
$
where $\eta$ is a random variable drawn from a two-sided geometric distribution. This distribution is parameterized to control the level of noise, typically governed by the privacy parameter $\varepsilon$.

\subsubsection{Geometric Distribution}
The two-sided geometric distribution is chosen due to its discrete nature and its ability to provide a privacy guarantee analogous to that of the Laplace mechanism for continuous data. The probability mass function (PMF) of the noise variable $\eta$ is given by

$
P(\eta = k) = \frac{1 - e^{-\varepsilon}}{1 + e^{-\varepsilon}} e^{-\varepsilon |k|}, \quad \text{for } k \in \mathbb{Z}.
$

This formulation ensures that the likelihood of larger deviations from the true count decreases exponentially with $|k|$, effectively managing the trade-off between privacy and accuracy.

\subsubsection{Privacy Guarantees}
The geometric mechanism satisfies $\varepsilon$-differential privacy for count queries with a known global sensitivity $\Delta f$. 
Typically, for count queries, the sensitivity is $ \Delta f = 1 $ (i.e., the addition or removal of a single individual can change the count by at most one). 
The mechanism ensures that the ratio of probabilities for any two neighboring datasets remains bounded by calibrating noise using the $\varepsilon$ parameter:

$
\frac{P(\tilde{f}(D) = r)}{P(\tilde{f}(D') = r)} \leq e^{\varepsilon},
$

where $D$ and $D'$ are neighboring datasets.

\textbf{Advantages}

\begin{itemize}
    \item \textbf{Discrete Data Suitability:} The mechanism is naturally suited for count queries and other discrete functions where the output is integer-valued.
    \item \textbf{Simplicity and Efficiency:} Its mathematical simplicity and ease of implementation make it attractive for practical applications in privacy-preserving data publishing.
    \item \textbf{Optimal Noise Distribution:} In many cases, the geometric distribution is optimal in minimizing the variance of the noise while still providing the required privacy guarantees.
\end{itemize}

\subsection{Exponential Mechanism}
The exponential mechanism is used to select an output from a set of possible outcomes by leveraging a scoring function. This mechanism is particularly useful for categorical data or when the output space is large, as it ensures that the probability of selecting an output is proportional to its score. The following sections detail the key aspects of the exponential mechanism.

Let $\mathcal{R}$ denote the set of possible outputs and $q(D, r)$ be a scoring function that assigns a utility value to each output $r \in \mathcal{R}$ given a dataset $D$. The exponential mechanism selects an output by assigning probabilities according to:

$
P(r \mid D) \propto \exp\left(\frac{\varepsilon \, q(D, r)}{2 \Delta q}\right),
$

where:
\begin{itemize}
    \item $\varepsilon$ is the privacy parameter,
    \item $\Delta q$ is the sensitivity of the scoring function, defined as the maximum change in $q$ when a single entry in the dataset is modified.
\end{itemize}

\subsubsection{Scoring Function and Selection}
The scoring function $q(D, r)$ is designed to reflect the quality or relevance of the output $r$ with respect to the dataset $D$. Higher scores indicate more desirable outcomes. The mechanism ensures that outputs with higher scores are selected with higher probability while still preserving privacy by exponentiating the score and normalizing over all possible outputs.

The selection probability is given by:

$
P(r \mid D) = \frac{\exp\left(\frac{\varepsilon \, q(D, r)}{2 \Delta q}\right)}{\sum_{r' \in \mathcal{R}} \exp\left(\frac{\varepsilon \, q(D, r')}{2 \Delta q}\right)}.
$

\subsubsection{Privacy Guarantees}
The exponential mechanism satisfies $\varepsilon$-differential privacy by ensuring that the relative probabilities of selecting any output from two neighboring datasets are bounded by $e^\varepsilon$. Specifically, for any two neighboring datasets $D$ and $D'$ and for any output $r \in \mathcal{R}$, the ratio of probabilities satisfies:

$
\frac{P(r \mid D)}{P(r \mid D')} \leq e^\varepsilon.
$

This means that the mechanism provides a strong privacy guarantee, ensuring that the output distribution does not significantly change when an individual's data is added or removed from the dataset. This guarantee is achieved by appropriately calibrating the score function's sensitivity $\Delta q$.

\textbf{Advantages}

\begin{itemize}
    \item \textbf{Flexibility for Categorical Data:} The exponential mechanism is ideal for scenarios where outputs are non-numeric or categorical.
    \item \textbf{Handling Large Output Spaces:} It is effective even when the set of possible outputs is large, as the mechanism scales by using a normalized probability distribution.
    \item \textbf{Utility Optimization:} the mechanism can be tailored to select outputs by leveraging a scoring function that maximizes utility while maintaining privacy.
\end{itemize}

\subsection{Randomized Response}
The randomized response technique provides privacy by ensuring that the probability of any particular response is similar, regardless of the respondent's true status. This makes it difficult to infer an individual's true answer from the observed response. We can balance the trade-off between privacy protection and the accuracy of the aggregate estimates by carefully choosing the probability $p$ for truthful responses, this can help achieve desired levels of privacy while still maintaining usefulness of data.

\textbf{Advantages}

\begin{itemize}
    \item \textbf{Enhanced Privacy:} Randomized response protects individual respondents from disclosure of sensitive information.
    \item \textbf{Simple Implementation:} The technique is straightforward to implement in surveys or questionnaires.
    \item \textbf{Accurate Aggregate Analysis:} Despite the noise introduced at the individual level, the method enables reliable estimation of overall population statistics.
\end{itemize}

This works by introducing randomness into the survey process. Each respondent is instructed to follow a randomizing procedure (for example, flipping a coin) before answering a sensitive question. Depending on the outcome of the randomization:

\begin{itemize}
    \item With a certain probability, the respondent answers truthfully.
    \item With the complementary probability, the respondent provides a predetermined random response.
\end{itemize}

This mechanism ensures that any single response does not reveal whether the respondent answered truthfully or not, thereby protecting individual privacy.

\subsubsection{Implementation}
Consider a survey question about a sensitive behavior. A typical implementation might involve the following steps:

\begin{figure}[htbp]
    \centering
    \includegraphics[width=0.35\textwidth]{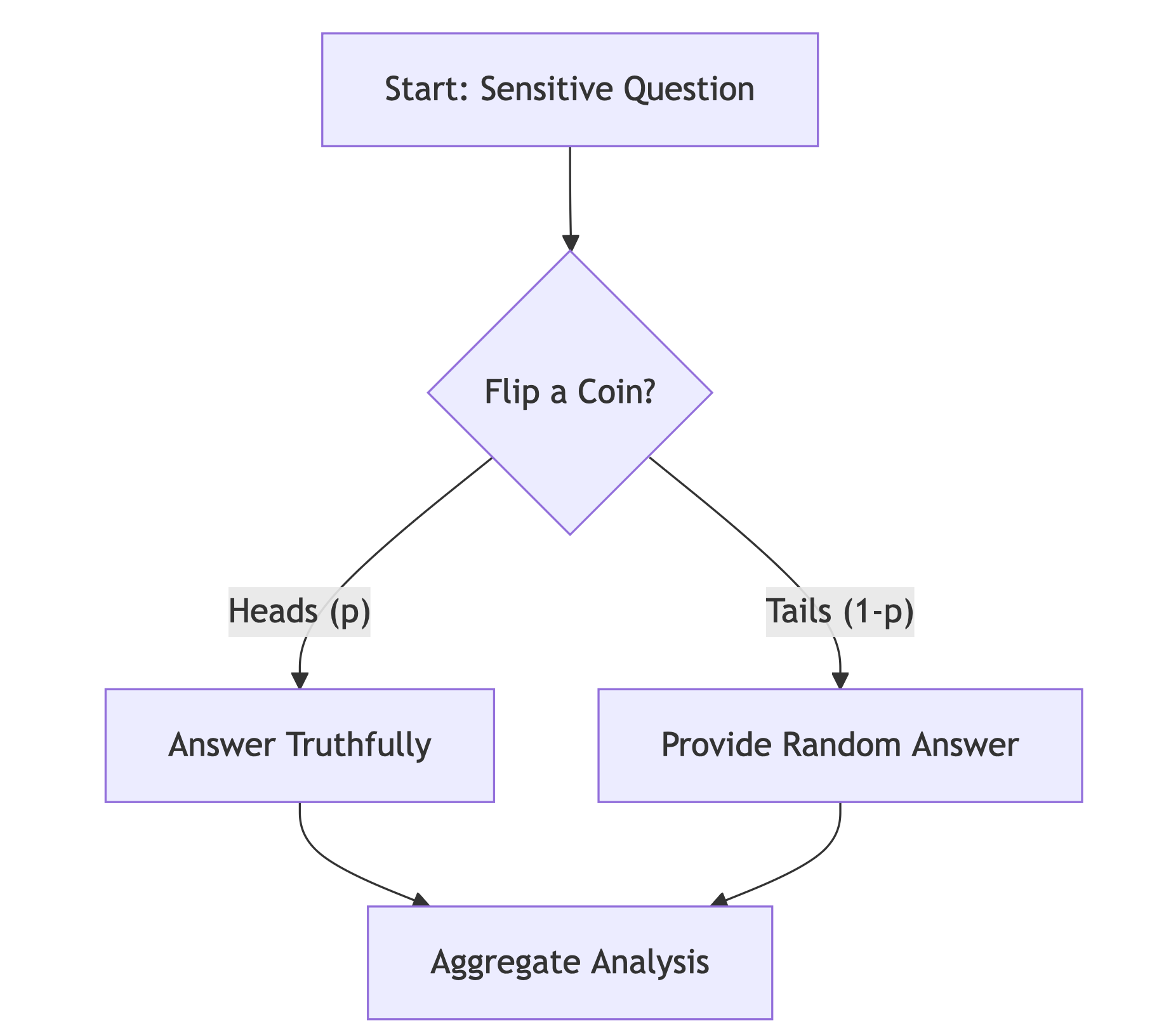}
    \caption{Randomized Response Implementation}
    \label{fig:randomized_response_example}
\end{figure}

This process introduces uncertainty in the individual responses while allowing the aggregate statistics to be estimated accurately.

\subsection{Laplace Mechanism}
The Laplace mechanism is a widely used technique in differential privacy that adds Laplace noise to numerical query outputs. It is particularly suitable for queries with known global sensitivity, ensuring that the added noise is calibrated to the sensitivity and the desired privacy level.

The Laplace mechanism is particularly effective for continuous data, where the addition of noise can be easily controlled to achieve the desired level of privacy.
\begin{figure}[htbp]
    \centering
    \includegraphics[width=0.8\textwidth]{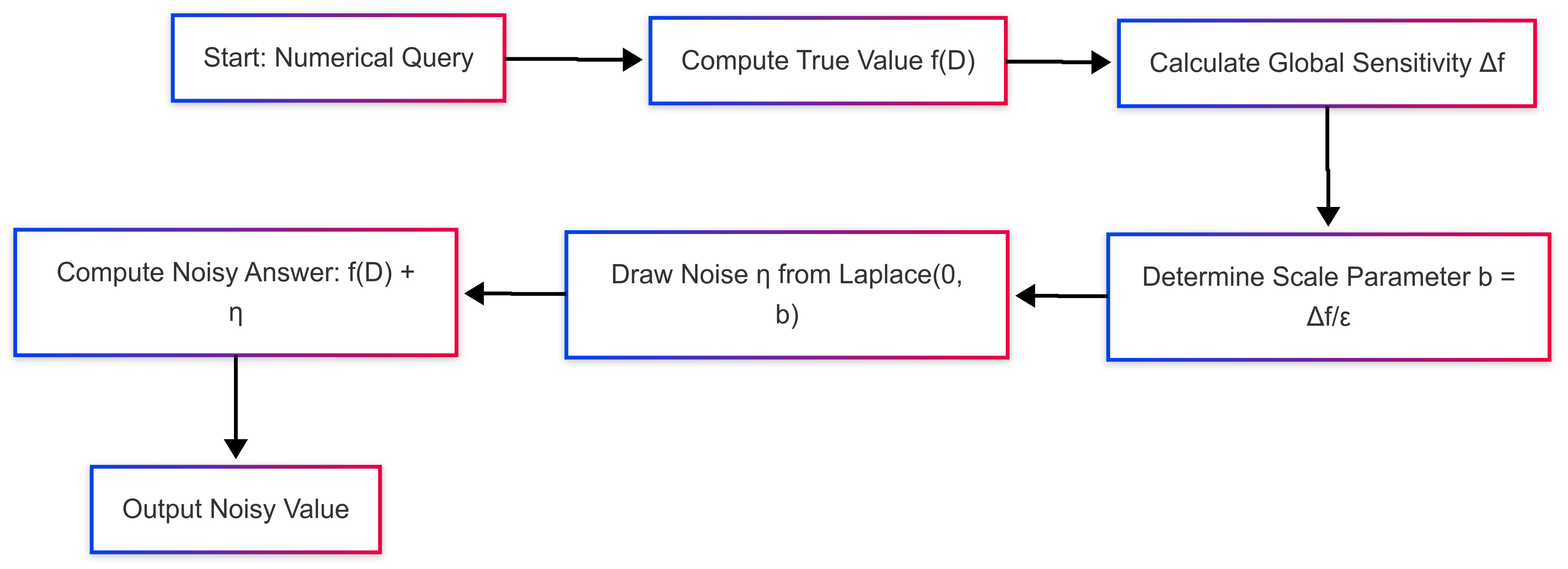}
    \caption{Laplace Mechanism Process}
    \label{fig:laplace_mechanism}
\end{figure}

For a query function $ f: \mathcal{D} \rightarrow \mathbb{R} $, the Laplace mechanism computes a noisy answer as follows:

$
\tilde{f}(D) = f(D) + \eta,
$

where $\eta$ is a random variable drawn from the Laplace distribution. The scale parameter $b$ of the Laplace distribution is set to 

$ b = \frac{\Delta f}{\varepsilon} $

where:

\begin{itemize}
    \item $\Delta f$ is the global sensitivity of the function $f$,
    \item $\varepsilon$ is the privacy parameter controlling the privacy-accuracy trade-off.
\end{itemize}

\subsubsection{Laplace Distribution}
The probability density function (PDF) of the Laplace distribution is given by:

$
\text{PDF}(\eta) = \frac{1}{2b} \exp\left(-\frac{|\eta|}{b}\right),
$

ensuring that the probability of larger noise values decreases exponentially. This calibration ensures that the mechanism satisfies $\varepsilon$-differential privacy.

\subsubsection{Privacy Guarantees}
The Laplace mechanism ensures that for any two neighboring datasets $D$ and $D'$, and for any output $r$, the ratio of probabilities satisfies:
$
\frac{P(\tilde{f}(D) = r)}{P(\tilde{f}(D') = r)} \leq 
e^{\varepsilon}.
$

This bound directly follows from the properties of the Laplace distribution, ensuring the desired level of privacy.

\textbf{Advantages}

\begin{itemize}
    \item \textbf{Suitability for Numerical Queries:} The Laplace mechanism is ideal for real-valued queries where the output is numerical.
    \item \textbf{Mathematical Simplicity:} Its straightforward formulation and analytical properties make it easy to implement and analyze.
    \item \textbf{Provable Privacy Guarantees:} The mechanism provides strong, provable privacy guarantees when the noise is appropriately calibrated.
\end{itemize}

\subsection{Gaussian Mechanism}
The Gaussian mechanism is employed in differential privacy by adding noise sampled from a Gaussian (normal) distribution to numerical query outputs. It is especially useful when aiming for $(\varepsilon, \delta)$-differential privacy guarantees, which are often required in iterative algorithms and settings where a small probability of failure is acceptable.

\begin{figure}[htbp]
    \centering
    \includegraphics[width=0.8\textwidth]{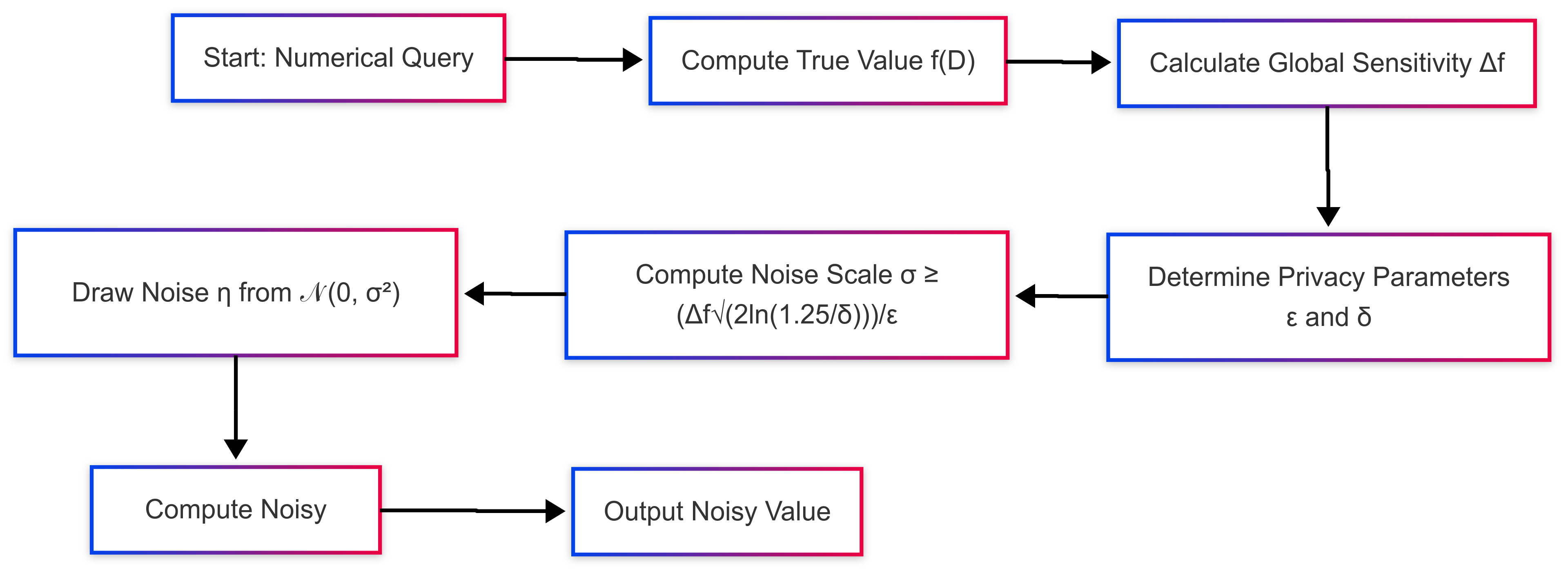}
    \caption{Gaussian Mechanism Process}
    \label{fig:gaussian_mechanism}
\end{figure}

For a query function $ f: \mathcal{D} \rightarrow \mathbb{R} $, the Gaussian mechanism outputs a noisy value defined by:
$
\tilde{f}(D) = f(D) + \eta,
$
where $\eta \sim \mathcal{N}(0, \sigma^2)$ is a Gaussian noise term with standard deviation $\sigma$ determined by the desired privacy parameters $\varepsilon$ and $\delta$, and the global sensitivity $\Delta f$ of the function.

\subsubsection{Determining the Noise Scale}
To achieve $(\varepsilon, \delta)$-differential privacy, the standard deviation $\sigma$ is typically calibrated as:

$
\sigma \geq \frac{\Delta f \sqrt{2 \ln(1.25/\delta)}}{\varepsilon}.
$

Here, $\Delta f$ represents the maximum change in the query function when a single entry in the dataset is modified, and the parameters $\varepsilon$ and $\delta$ control the privacy-accuracy trade-off.

\subsubsection{Privacy Guarantees}
The Gaussian mechanism ensures that the ratio of the probabilities of any output given two neighboring datasets $D$ and $D'$ is bounded, providing $(\varepsilon, \delta)$-differential privacy. This guarantee is derived from the properties of the Gaussian distribution and the careful calibration of the noise level.

\textbf{Advantages}

\begin{itemize}
    \item \textbf{Applicability to Iterative Algorithms:} The Gaussian mechanism is particularly effective in settings where multiple rounds of queries or iterative computations are performed.
    \item \textbf{Handling Complex Queries:} It is well-suited for complex numerical queries where the output is continuous.
    \item \textbf{Fine-Tuning Privacy Guarantees:} The mechanism offers flexibility in balancing privacy and utility by adjusting both $\varepsilon$ and $\delta$, allowing for a more nuanced approach to privacy protection.
    \item \textbf{Robustness to Outliers:} The Gaussian mechanism is less sensitive to extreme values compared to the Laplace mechanism, making it suitable for datasets with heavy-tailed distributions.
\end{itemize}

\subsection{Advanced PII Transformation}
Advanced PII Transformation is a context-sensitive technique that aims to identify personal identifiable information (PII) within data and transform it into plausible but fictitious values, referred to as Faux-PII. Unlike simple redaction, this method preserves the original format and plausibility of the data, thereby maintaining higher utility for downstream applications. The technique is inspired by principles outlined in “Life of PII” \cite{deshmukh2023lifepiipii} and “TableGuard” \cite{tableguard}.

\begin{figure}[htbp]
    
    \centering
    \includegraphics[width=0.8\textwidth]{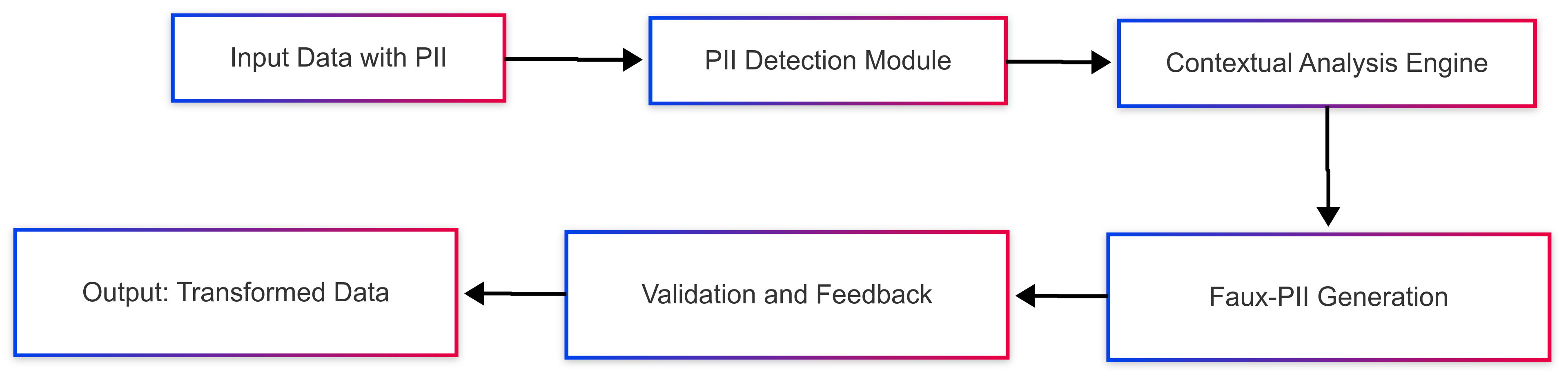}
    \caption{Advanced PII Transformation Process}
    \label{fig:advanced_pii_transformation}
\end{figure}

The core idea behind Advanced PII Transformation is to obfuscate sensitive data elements while retaining their semantic and contextual properties. This is achieved by:
\begin{itemize}
    \item \textbf{Contextual Identification:} Utilizing natural language processing (NLP) and named entity recognition (NER) techniques to accurately identify PII within text or structured data.
    \item \textbf{Faux-PII Generation:} Transforming the identified PII into realistic but fictitious values that mimic the original format and context.
    \item \textbf{Utility Preservation:} Maintaining the overall data structure and semantic integrity, which allows for effective data analysis and minimizes information loss.
\end{itemize}

\subsubsection{Key Components}
\begin{itemize}
    \item \textbf{PII Detection Module:} Applies advanced NLP techniques (e.g., NER, PoS tagging) to locate and classify PII elements.
    \item \textbf{Contextual Analysis Engine:} Evaluates the surrounding context to ensure that the transformation preserves both the format and plausibility of the data.
    \item \textbf{Transformation Algorithm:} Maps original PII values to synthetic counterparts (Faux-PII) that are realistic yet non-identifiable.
    \item \textbf{Validation and Feedback:} Measures potential information loss and refines the transformation process to balance privacy with data utility.
\end{itemize}

The Advanced PII Transformation process can be visualized as a pipeline, where each component interacts to ensure that the final output is both privacy-preserving and analytically useful. The diagram in Figure \ref{fig:advanced_pii_transformation} illustrates this flow, highlighting the key components and their interactions.

\textbf{Advantages}

\begin{itemize}
    \item \textbf{Enhanced Data Utility:} Preservation of structural and semantic meaning, the transformed data remains useful for analytical tasks.
    \item \textbf{Context-Sensitive Protection:} The transformation considers the surrounding context, reducing the risk of cognitive dissonance compared to simple redaction.
    \item \textbf{Flexible Application:} Suitable for both structured and unstructured data, making it valuable for databases, documents, and real-time data streams.
\end{itemize}

\subsection{Generative Models}
Generative models, such as Generative Adversarial Networks (GANs) and Variational Autoencoders (VAEs), learn the underlying distribution of real data and can generate entirely new, artificial data points that closely resemble the original. When combined with differential privacy techniques (e.g., DP-SGD), these models enable the creation of synthetic data with formal privacy guarantees. This approach not only preserves data utility but also protects sensitive information.

\begin{figure}[htbp]
    \centering
    \includegraphics[width=0.8\textwidth]{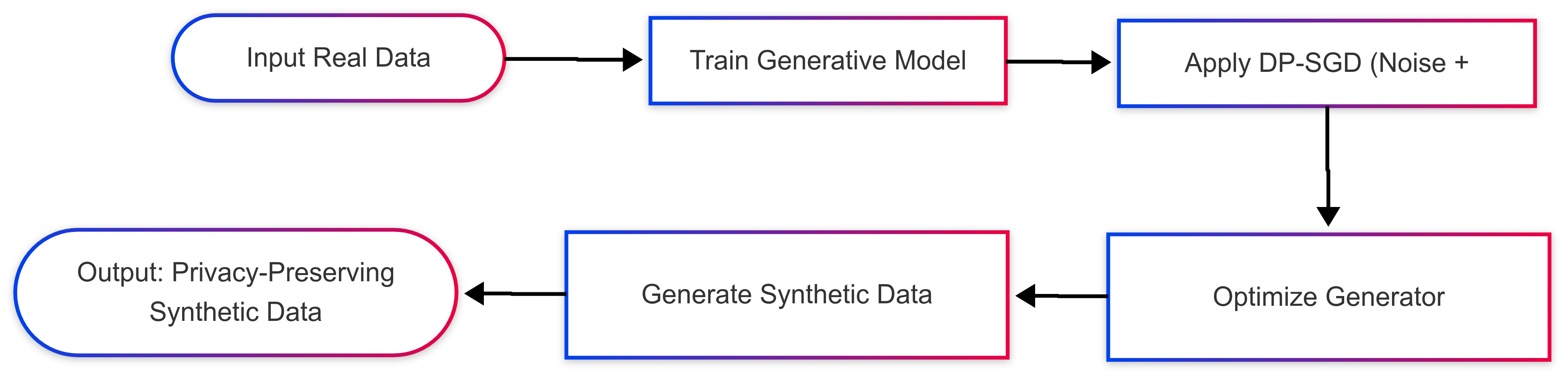}
    \caption{Generative Modeling Process with Differential Privacy}
    \label{fig:generative_modeling_process}
\end{figure}

\subsubsection{Mechanism Overview}
Generative models are designed to model complex data distributions. They typically involve two main components:
\begin{itemize}
    \item \textbf{Generator:} Learns to produce synthetic data samples from a latent space.
    \item \textbf{Discriminator (or Encoder in VAEs):} Evaluates the quality of the generated samples by comparing them to real data.
\end{itemize}
The generator captures the underlying data distribution, allowing it to create new samples that are statistically similar to the original dataset.

\subsubsection{Differential Privacy in Model Training}
Incorporating differential privacy into the training process, such as via Differentially Private Stochastic Gradient Descent (DP-SGD), provides formal privacy guarantees. DP-SGD works by:
\begin{itemize}
    \item Adding carefully calibrated noise to the gradients during training.
    \item Clipping gradients to ensure that the contribution of any single data point is bounded.
\end{itemize}
This approach limits the influence of individual data records, ensuring that the generated synthetic data does not reveal sensitive information about any single entry in the original dataset.

\textbf{Advantages}

\begin{itemize}
    \item \textbf{High Data Utility:} Generated data retains statistical properties of the original dataset, making it useful for downstream tasks.
    \item \textbf{Privacy Protection:} Combining generative models with differential privacy mechanisms provides formal privacy guarantees.
    \item \textbf{Flexibility:} Applicable to a wide range of data types, including images, text, and structured data.
\end{itemize}

\section{Challenges in Traditional Approaches}
Traditional privacy-preserving methods, despite offering privacy benefits, often struggle to strike the right balance between utility and privacy. In this section, we detail the key challenges associated with these approaches, focusing on noise calibration, binning bias, and the distortion of data distributions through perturbation techniques.

\begin{figure}[htbp]
    \centering
    \includegraphics[width=0.8\textwidth]{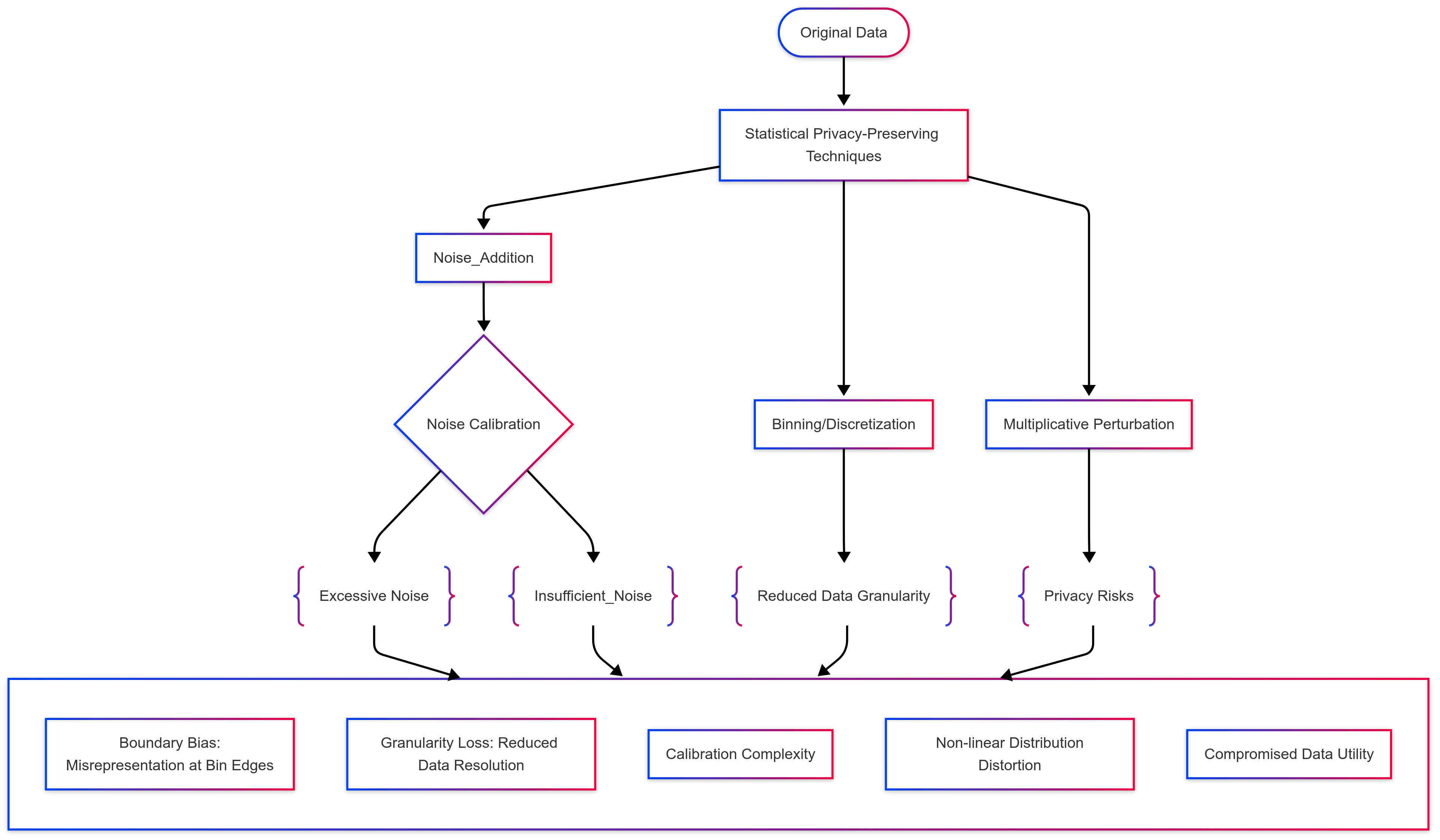}
    \caption{Challenges in Traditional Privacy-Preserving Approaches}
    \label{fig:challenges_in_traditional_approaches}
\end{figure}

\subsubsection{Utility vs. Privacy Trade-off}
One of the fundamental challenges in privacy-preserving data publishing is achieving a balance between data utility and privacy protection. Methods that inject excessive noise can render datasets unsuitable for analytical tasks, whereas insufficient noise can leave sensitive information vulnerable. Mathematically, if a query function $ f $ is perturbed by noise $ \eta $ (e.g., using the Laplace mechanism), the noisy output is:
$
\tilde{f}(D) = f(D) + \eta,
$
where the noise scale $ b $ is typically set to $ b = \Delta f/\varepsilon $. A lower $\varepsilon$ (stronger privacy) results in higher $ b $, which can significantly reduce the utility of the data.

\subsubsection{Excessive Noise and Data Utility}
Excessive noise addition is a common issue that leads to high variance in the output data, making it difficult to extract meaningful insights. For example:
\begin{itemize}
    \item \textbf{Analytical Degradation:} Statistical metrics such as means, variances, and correlations may deviate substantially from their true values.
    \item \textbf{Model Performance:} Machine learning models trained on noisy data may exhibit poor predictive performance due to the distortion of underlying patterns.
\end{itemize}

\subsubsection{Insufficient Noise and Privacy Leakage}
Insufficient noise fails to obscure the sensitive information effectively. This can occur when the noise calibration does not account for the true sensitivity of the data, potentially exposing individual records. This challenge is particularly critical in high-dimensional datasets where the cumulative effect of small privacy leaks can be significant.

\subsubsection{Bias from Binning}
Binning is often employed to reduce the sensitivity of continuous variables by grouping data into intervals. However, this approach introduces its own set of challenges:
\begin{itemize}
    \item \textbf{Boundary Effects:} The choice of bin boundaries can introduce systematic biases. Data points that lie near the edges of bins might be misrepresented, leading to distorted statistical properties.
    \item \textbf{Loss of Granularity:} Aggregating data into bins reduces the level of detail, which may negatively impact analyses that rely on fine-grained information.
\end{itemize}

\subsubsection{Perturbation Techniques and Calibration}
Perturbation methods, such as multiplicative noise, require careful calibration to avoid significant distortion of the original data distribution. Key challenges include:
\begin{itemize}
    \item \textbf{Non-linear Distortion:} Multiplicative perturbations can alter the underlying distribution in a non-linear fashion, especially when the original data contains outliers.
    \item \textbf{Calibration Complexity:} Determining the appropriate noise level often involves complex trade-offs and domain-specific considerations, making it difficult to generalize across different datasets.
\end{itemize}

\section{Use Cases in BFSI}
The BFSI sector is particularly well-suited for the application of synthetic data generation and advanced perturbation techniques due to its data-rich environment and the critical need for robust analytics. Financial institutions are increasingly turning to machine learning and AI to drive insights, improve customer experiences, and enhance operational efficiency. 

However, the BFSI sector presents unique challenges and opportunities due to its sensitive nature of data and its regulatory landscape, the complexity of financial products, and the need for high-stakes decision-making based on data-driven insights.

Industries such as Healthcare, Retail, and Telecommunications have also seen significant advancements in privacy-preserving data management. 

\subsection{Fraud Detection}
Synthetic data can be used to train machine learning models for fraud detection without exposing sensitive customer information. Generating realistic transaction patterns, financial institutions can improve their models' performance while ensuring compliance with privacy regulations.

\subsection{Risk Assessment}
Synthetic data can be used to create diverse scenarios for risk assessment models, allowing financial institutions to evaluate potential risks without relying on real customer data. This can help in stress testing and scenario analysis while maintaining customer privacy.

\subsection{Customer Segmentation}
Synthetic data can be used to create customer profiles for segmentation analysis, enabling financial institutions to tailor their marketing strategies without exposing sensitive customer information and improve their targeting and personalization efforts.

\subsection{Regulatory Compliance}
Synthetic data can help financial institutions comply with regulations like GDPR and CCPA by providing a privacy-preserving alternative to real customer data to reduce the risk of data breaches and ensure compliance with privacy regulations.

\subsection{Operational Efficiency}
Synthetic data can be used to streamline data processing and analytics workflows, reducing the time and resources required for data preparation, testing and validation. Financial institutions can improve their operational efficiency while maintaining data privacy.

\subsection{Data Augmentation}
Synthetic data can be used to augment existing datasets, especially in cases where real data is scarce or imbalanced. Using statistical techniques we can generate additional samples to offset imbalances, financial institutions can improve the performance of their machine learning models and enhance their analytical capabilities without compromising customer privacy.

\subsection{Data Sharing and Collaboration}
Synthetic data can facilitate secure data sharing and collaboration between financial institutions, enabling them to share insights and analytics without exposing sensitive customer information.

\subsection{Data-Driven Decision Making}
Synthetic data can support data-driven decision-making processes by providing privacy-preserving datasets for analytics and reporting while ensuring compliance with privacy regulations.

\subsection{Data-Driven Customer Experience}
Synthetic data can be used to enhance customer experience by enabling personalized recommendations and targeted marketing strategies without exposing sensitive customer information.

\section{Conclusion}

\begin{itemize}
    \item While the Laplace mechanism is widely used for continuous queries, its direct application to discrete queries can lead to rounding issues. The geometric mechanism circumvents this problem by directly generating discrete noise, making it a preferred option when the output domain is inherently discrete.
    \item The exponential mechanism is particularly useful for selecting outputs from large or complex output spaces, where a scoring function can be employed to prioritize high-utility outputs. This mechanism is flexible and can be adapted to various applications, including categorical data and complex decision-making scenarios.
    \item Unlike mechanisms that add noise to numerical outputs (e.g., the Laplace or geometric mechanisms), the exponential mechanism is better suited for non-numeric outputs. Its probabilistic selection process ensures that high-utility outputs are favored, making it particularly useful when dealing with complex or large output spaces.
    \item Randomized response is a well-established technique for collecting sensitive data while preserving privacy. It is particularly effective in survey settings where respondents may be reluctant to disclose sensitive information. This method allows for accurate aggregate analysis without revealing individual responses.
    \item The Laplace mechanism is a widely used approach for adding noise to numerical queries, ensuring privacy while maintaining utility. It is particularly effective for continuous data and is often employed in various applications, including statistical analysis and machine learning.
    \item The Gaussian mechanism is an extension of the Laplace mechanism, providing privacy guarantees in scenarios where the output space is continuous and the sensitivity of the query function is known. It is particularly useful for iterative algorithms and can be applied to a wide range of applications, including machine learning and data analysis.
    \item Advanced PII Transformation offers significant benefits over traditional redaction:
    \begin{itemize}
        \item \textbf{Format Preservation:} Unlike redaction, which removes data, this approach maintains the original format.
        \item \textbf{Realistic Data Substitution:} The generated Faux-PII is designed to be plausible, which is critical for applications such as testing, analytics, or feeding data to machine learning models.
        \item \textbf{Reduced Information Loss:} Transforming PII rather than eliminating it minimizes the loss of valuable context.
    \end{itemize}

\end{itemize}

\section{Future areas of research}
\begin{itemize}
    \item \textbf{Adaptive Mechanisms:} Future research could focus on developing further adaptive mechanisms that dynamically adjust noise levels based on the sensitivity of the data and the specific query being executed.
    \item \textbf{Integration with Machine Learning:} Exploring how differential privacy can be seamlessly integrated into machine learning pipelines, particularly in federated learning settings, could yield significant advancements in privacy-preserving AI \cite{wang}.
    \item \textbf{Real-World Applications:} Continued exploration of real-world applications, particularly in sensitive domains like healthcare and finance, will help refine techniques and demonstrate their practical utility.
    \item \textbf{User-Centric Approaches:} Research into user-centric approaches that allow individuals to control their data privacy settings while still enabling effective data sharing and analysis could lead to more robust privacy solutions.
    \item \textbf{Ethical Considerations:} Investigating the ethical implications of synthetic data generation and privacy-preserving techniques will be crucial to ensure responsible use of these technologies.
    \item \textbf{CFE and Data Provenance:} Future research could explore the integration of CFE techniques with data provenance to enhance the traceability and accountability of synthetic data generation processes.
\end{itemize}

\section{Acronyms}

GAN = Generative Adversarial Networks

VAE = Variational Autoencoder

DP = Differential Privacy

PII = Personally Identifiable Information

BFSI = Banking, Financial Services, and Insurance

GDPR = General Data Protection Regulation

CCPA = California Consumer Privacy Act

LDP = Local Differential Privacy

CDP = Central Differential Privacy

DP-SGD = Differentially Private Stochastic Gradient Descent

CFE = Counterfactual Fairness Evaluation

\bibliographystyle{unsrt}
\bibliography{references}

\end{document}